\documentclass[epjST]{svjour}
\usepackage{graphics}
\usepackage{float}
\usepackage{hyperref}       
\usepackage{url}
\newcommand{\be}{\begin{equation}}
\newcommand{\ee}{\end{equation}}
\newcommand{\e}{\varepsilon}
\begin{document}
\title{Minimal Chimera States in Phase-Lag Coupled Mechanical Oscillators}
\author{P. Ebrahimzadeh\inst{1}\fnmsep\thanks{\email{p.ebrahimzadeh@fz-juelich.de}} \and M. Schiek \inst{1}\fnmsep\thanks{\email{m.schiek@fz-juelich.de}}
 \and P. Jaros  \inst{2}\fnmsep\thanks{\email{patrycja.kuzma@p.lodz.pl}} \and T. Kapitaniak \inst{2}\fnmsep\thanks{\email{tomaszka@p.lodz.pl}} \and S. van Waasen \inst{1}\fnmsep\thanks{\email{s.van.waasen@fz-juelich.de}} \and Y. Maistrenko \inst{1,2,3}\fnmsep\thanks{\email{y.maistrenko@biomed.kiev.ua}}}
\institute{Forschungszentrum J\"{u}lich GmbH, ZEA-2: Electronics Systems, 52428 J\"{u}lich, Germany \and Division of Dynamics, Technical University of Lodz, Stefanowskiego 1/15, 90-924 Lodz, Poland \and Institute of Mathematics and Center for Medical and Biotechnical Research, National Academy of Science of Ukraine, Tereshchenkivska St. 3, 01030 Kyiv, Ukraine}
\abstract{
We obtain experimental chimera states in the minimal network of three identical mechanical oscillators (metronomes), by introducing phase-lagged all-to-all coupling. For this, we have developed a real-time model-in-the-loop coupling mechanism that allows for flexible and online change of coupling topology, strength and phase-lag. The chimera states manifest themselves as a mismatch of average frequency between two synchronous and one desynchronized oscillator. We find this kind of striking “chimeric” behavior is robust in a wide parameter region. At other parameters, however, chimera state can lose stability and the system behavior manifests itself as a heteroclinic switching between three saddle-type chimeras. Our experimental observations are in a qualitative agreement with the model simulation.
} 
\maketitle
\section{Introduction}
\label{intro}
Chimera states, discovered at the edge of millennium by Kuramoto and Battogtohk \cite{kuramoto2002coexistence,PhysRevLett.93.174102}, are spatiotemporal patterns emerging in high-dimensional oscillatory networks  as a coexistence of coherent and incoherent groups of oscillators~\cite{PhysRevLett.101.084103,LAING20091569,PhysRevLett.104.044101,motter2010spontaneous,PhysRevE.81.065201,PhysRevLett.106.234102,doi:10.1142/S0218127414400148,PhysRevLett.112.144103,kemeth2016classification}. Ever since, these remarkable counter-intuitive patterns have been a subject to intensive theoretical investigations. Beginning from 2012, chimera states have been observed experimentally in optics and opto-electronics~\cite{PhysRevLett.111.054103,article}, chemical and electrochemical~\cite{article2,doi:10.1063/1.4858996,article3}, and mechanical oscillator systems~\cite{Martens10563,Kapitaniak2014,Wojewoda2016}. They have been discussed to play role in the complex dynamics, e.g. neural networks~\cite{doi:10.1063/1.5009812,Hizanidis2016}, power grids~\cite{Motter2013,Pecora2014} and social systems~\cite{GONZALEZAVELLA201424}. One can find a detailed review on the role of chimera states in neuronal networks in~\cite{MAJHI2019100}. To this end, chimera states attract a growing attention, as illustrated in recent review papers by Pannagio and Abrams [2015], Sch\"{o}ll [2017] and, Omelchenko [2018]. 

Recently, it was reported theoretically that chimera states can also exist in small networks of coupled oscillators. These so-called {\it weak} chimera states, defined by Ashwin and Burylko in 2015~\cite{doi:10.1063/1.4905197}, refer to trajectories in which at least one oscillator drifts in phase with respect to a frequency-synchronized group. This behavior is then manifested in the difference between respective average frequencies (i.e. Poincare winding numbers). In the context of minimal networks, the definition of weak chimera states is equivalent to so-called solitary states \cite{PhysRevE.91.022907,doi:10.1063/1.5019792,doi:10.1063/1.5061819}. The other notable behavior in small networks is the heteroclinic switching between saddle chimera states~\cite{PhysRevX.10.011044,articleAshwin,PhysRevE.97.050201}. This type behavior arises after the existing chimeras lose stability transforming into saddles. In the $N=3$ case, due to the $S_3$ symmetry, the switching is then performed between three saddle-type chimeras~\cite{PhysRevE.95.010203}. 

\begin{figure}
  \resizebox{1.\columnwidth}{!}{
  \includegraphics{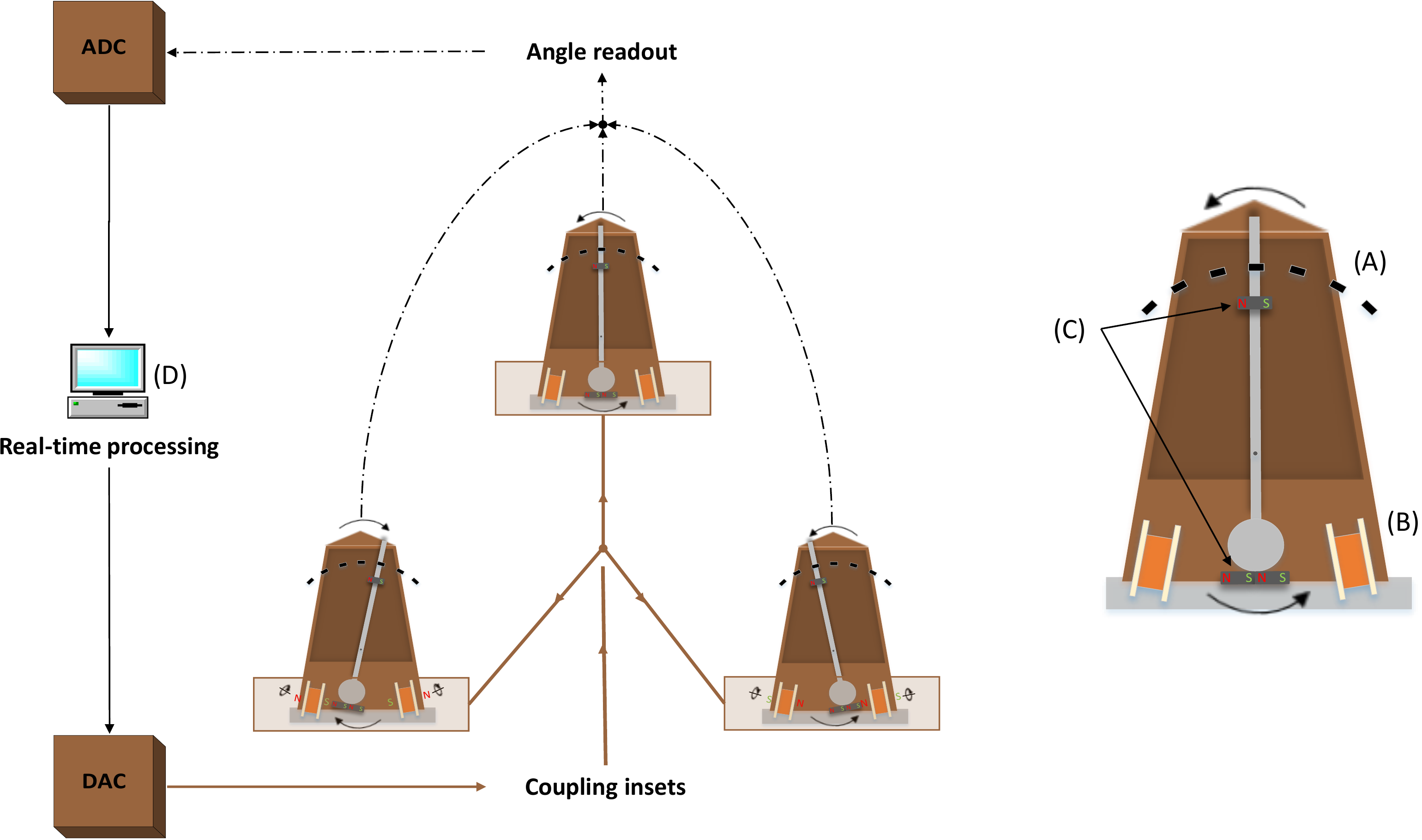}}
  \caption{Experimental setup configuration. The three metronomes are equipped with an array of magnetic sensors (Hall sensors, A) for contactless measuring of the rod’s position and two coils (B) for applying magnetic force onto the counter weight. To increase magnetic signal and force Neodymium magnets are attached to rod and counterweight (C). Hall sensor signals (dashed line) are digitized in the ADC unit. In real-time processing unit (D), the angles of all metronomes are reconstructed based on the ADC data and the coupling terms are calculated according to Eq.~\ref{K1}. In DAC unit the coupling terms are converted to electrical currents (solid brown lines) passing through the coils and generating magnetic forces. (See Appendix)}
  \label{fig1}
\end{figure}

In spite of intensive theoretical investigations, experimental realization of chimera states in minimal networks is still challenging. In this article, we report the experimental observation of the chimera dynamics in a mechanical system of three coupled metronomes with variable phase-lag and coupling strength. For the purpose, we have built a setup in which the coupling between the metronomes is organized in real-time via a computer, as illustrated in Fig.~\ref{fig1}. Due to the flexibility of such approach, we may electronically assign any desired coupling scheme with the coupling magnitude $\mu_{i,j}$ and phase-lag $\alpha_{i,j}$, then vary the parameters in a wide range. Our setup with three all-to-all coupled metronomes allows to observe stable anti-phase chimera states,  i.e. those in which two oscillators are synchronized in anti-phase and the third one is frequency deviating, as well as heteroclinic switching between the chimeras, both capturing wide regions of the parameter space.  
\section{Experimental Setup and the Model}
\label{model}
The setup is shown schematically in Fig.~\ref{fig1}. It contains three standard mechanical metronomes, each of which is a pendulum of length $l$, mass $m$ and moment of inertia $B$. The position of each metronome’s rod is registered by an arrangement of 7 Hall sensors. For creation of a strong magnetic signal, a small Neodymium magnet is mounted on the rod. Based on the signals of the Hall sensors, the rod angle $\theta_i$ is then calculated by interpolation of a look-up table which is created in advance per calibration procedure. The coupling force is generated by an electrical current passing through the coils attached on both sides of the counter weight, the current being proportional to the respective coupling term. This method of implementation allows for flexible definition and online variation of coupling strength $\mu$ and phase-lag $\alpha$. Moreover, this real-time computation-implementation coupling mechanism ensures a fixed response time well below $1 ms$. 

The motion of each pendulum is damped by the linear damper characterized by the coefficient $\e$. To oppose this damping, the internal escapement mechanism of the metronome generates an excitation torque $M_D$ in the threshold region $(-\gamma_N, \gamma_N)$: in the first stage when $0 < \theta_i < \gamma_N$ the escapement mechanism activates and $M_D = M_N$, and when $\theta_i < 0$ it is deactivated $M_D = 0$. Similarly, in the second stage for $ -\gamma_N < \theta_i < 0$ activation works as  $M_D = - M_N$, and for $\theta_i > 0$  $M_D = 0$.  

An important characteristic of each metronome is the multistability of its dynamics. Beyond the standard oscillatory regime given by a limit cycle, there are two silent states corresponding to stable fixed points $A^+, A^-$ $(\pm \sin^{-1}M_N/mgl, 0)$. Depending on the initial angle $\theta_{i,0}$ and the initial velocity  $\dot{\theta}_{i,0}$, each of the three regimes can be realized. If $|\theta_{i,0}|>\gamma_N$  the dynamics is always oscillatory, otherwise it drops to the silence as soon as phase point ($\theta_{i,0}$, $\dot{\theta}_{i,0}$) lies inside the ellipsoid $\theta_{i,0}/{{\gamma_N}^2}+ \dot{\theta}_{i,0}/\omega{{\gamma_N}^2}=1$, Ref.~\cite{Wojewoda2016}.
\begin{figure}
\resizebox{1.\columnwidth}{!}{
 \includegraphics{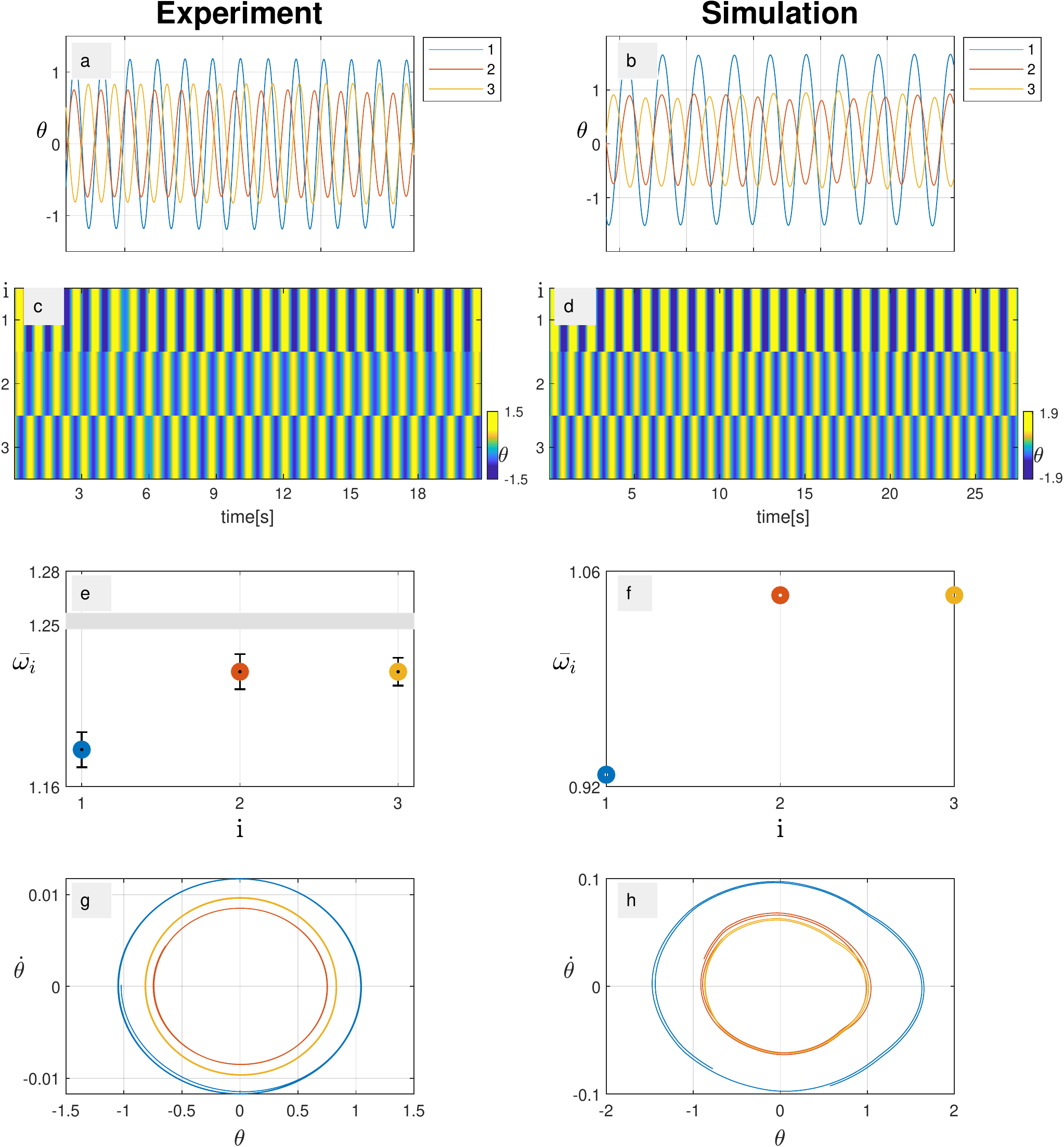} }
\caption{(Color Online) Chimera state in coupled metronome system at parameters $\alpha = 1.5$ and $\mu = 9 \times 10^{-4}$, for experiment (left) and simulation (right). (a),(b) Angle time series of metronomes showing the phase-locked metronomes no. 2 and 3, while no.1 changes its state, i.e. drifts in phase. (c),(d) color map representation of (a) and (b), respectively. (e),(f) frequency profiles showing the difference between phase-locked and drifting metronomes indicating system to be in {\it weak} chimera state, see Ref.~\cite{doi:10.1063/1.5009812}. The eigenfrequency of metronomes reside in the grey region of (g) where their difference is calibrated less than $1\%$. (g),(h) phase portraits of approximately two periods of time interval of (a) and (b). The video of the phases portraits of the whole time interval can be found at [\url{https://fz-juelich.sciebo.de/s/w45BvSge3cAIYH6}]. }
\label{fig2}
\end{figure}
The dynamics of the three-metronome system in Fig.~\ref{fig1} is modeled using the equations of motion derived from the principles of classical mechanics. The angle dynamics is then described by  the Kuramoto model with inertia
\begin{equation}
\label{K1}
B \ddot{\theta_i} + \e \dot{\theta_i} + m g l \sin\theta_i = M_D + H(\theta_i-\theta^{\ast}) \frac{\mu}{3} \sum_{i=1}^{3} \sin(\theta_j - \theta_i - \alpha), 
\end{equation}
in which the sinusoidal coupling term is controlled by the Heaviside-like step function $H(\theta-\theta^{\ast}$) according to the position of the coils $\theta^{\ast}\approx30^{\circ}$. The experiments and numerical simulation consisted of varying the coupling strength $\mu$ and phase-lag $\alpha$. 
\section{Results}
\label{Results}
Our main result consists in the experimental observation of chimera states for the network of three all-to-all coupled metronomes. In the ''chimeric'' behavior, two of the metronomes are frequency synchronized in anti-phase (normally with a breathing) but the third one exhibits a different average frequency, satisfying thereby the definition of {\it weak} chimera state~\cite{doi:10.1063/1.4905197}. Typical example of the experimental chimera state is presented in Fig.~\ref{fig2}(left). The state is obtained for practically identical metronomes (natural frequencies are calibrated for less than $1\%$ deviations), and for  fully symmetric coupling strength controlled by a computer. We find that this kind chimeras arise at large enough values of the phase-lag parameter $\alpha$,  beginning from $\alpha\approx0.7$, and they preserve with further increase of $\alpha$, even beyond $\pi/2$. Robustness of the experimental chimeras is additionally supported by the fact that they originate from arbitrary assigned initial conditions (as long as all three metronomes are in the oscillatory regime). Numerical simulations of Eq.~(\ref{K1}) demonstrate qualitative agreement with the experiment, see Fig.~\ref{fig2}(right).

Full parameter region for the chimera state illustrated in Fig.~\ref{fig2} is plotted  schematically in Fig.~\ref{fig3}; simulation details can be found in Appendix. The example illustrated in Fig.~\ref{fig2} corresponds to parameter point $A$. We observed that chimera preserves its properties in the whole region schown in grey. 
\begin{figure}
\resizebox{1.\columnwidth}{!}{
 \includegraphics{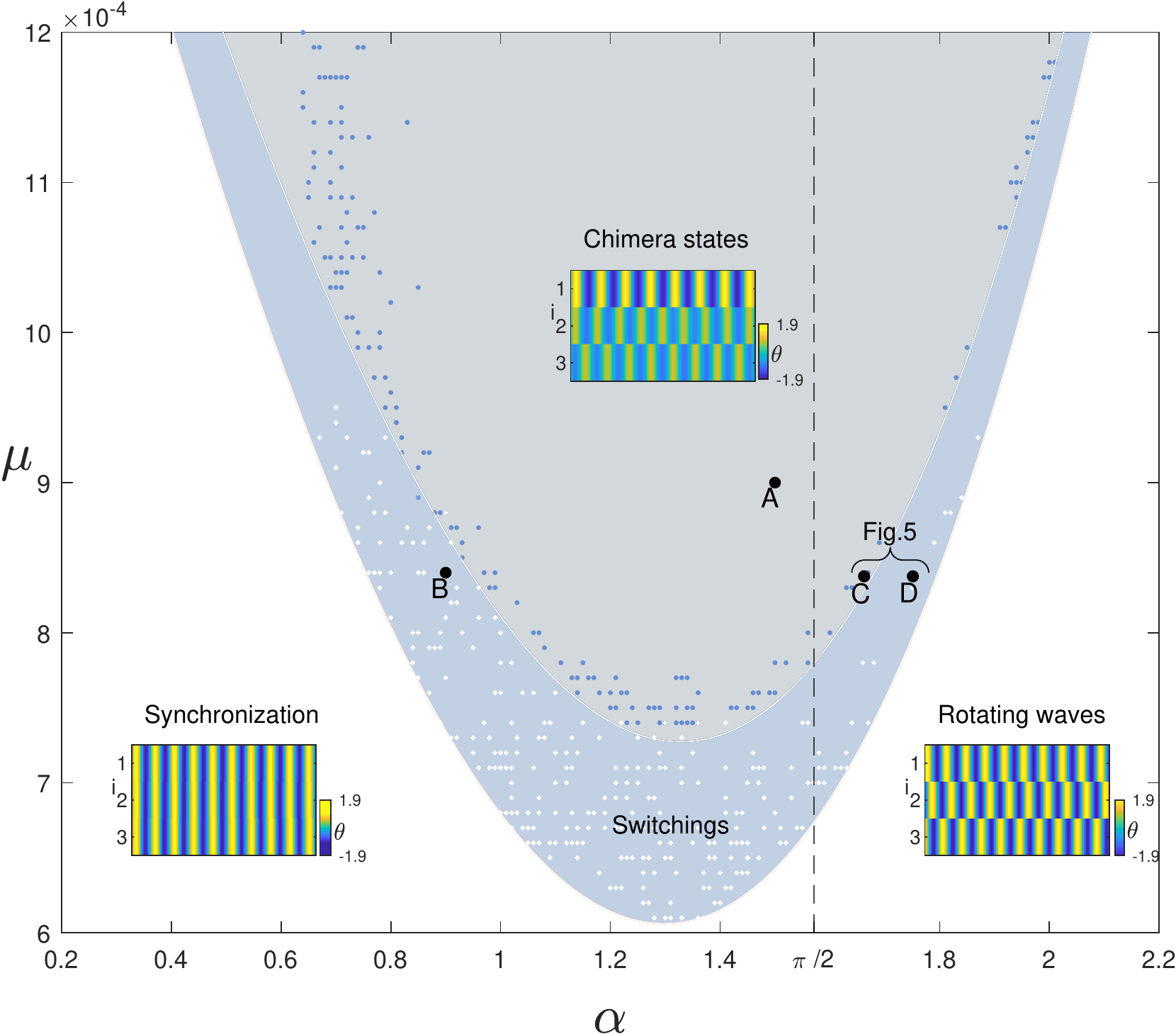} }
\caption{ Parameter region of dependence of different spatiotemporal patterns on parameters $\alpha$ and $\mu$. The white data points in the {\it Switchings} region indicate coexistence of the synchronous state (for $\alpha < \pi/2$) and rotating waves (for $\alpha > \pi /2$) with heteroclinic switchings. The border between {\it Chimera states} and {\it Switchings} are not clear cut, heteroclinic switchings infiltrated the chimera region are shown as blue dots. The sample points $A$ and $B$ are presented in Figs.~\ref{fig2} and \ref{fig4a}, respectively.}
\label{fig3}
\end{figure}
Moving left from the sample parameter point $A$ by decreasing phase-lag $\alpha$, chimera state loses its stability transforming into a saddle chimera. The system behavior represents the so-called {\it heteroclinic switching} between three saddle chimera states (existing in this case due to the symmetry of the problem). Heteroclinic switching is the second characteristic regime observed in our experiment. Parameter region for it is shown in Fig.~\ref{fig3} in light blue color. An example of the switching behavior, point $B$, is presented in Fig.~\ref{fig4a}. With further decrease of the phase-lag parameter $\alpha$, the metronomes synchronize. Alternatively, moving right from sample point $A$ with increase of $\alpha$ beyond $\pi/2$, heteroclinic switchings coexist with rotating waves. The basins of attraction for chimera states and heteroclinic switchings show riddling-like pattern, see Fig.~\ref{fig5}. These states are very sensitive to initial conditions. Experimentally, we observed that any small perturbation can result in ''loss'' of chimera state or irregular switches between saddle chimeras, which mostly falls into a rotating wave. With further increase of $\alpha$ beyond the right boundary of this region, collective metronome dynamics in experiment and simulation demonstrate rotating waves.  
\begin{figure}
\resizebox{1.\columnwidth}{!}{
 \includegraphics{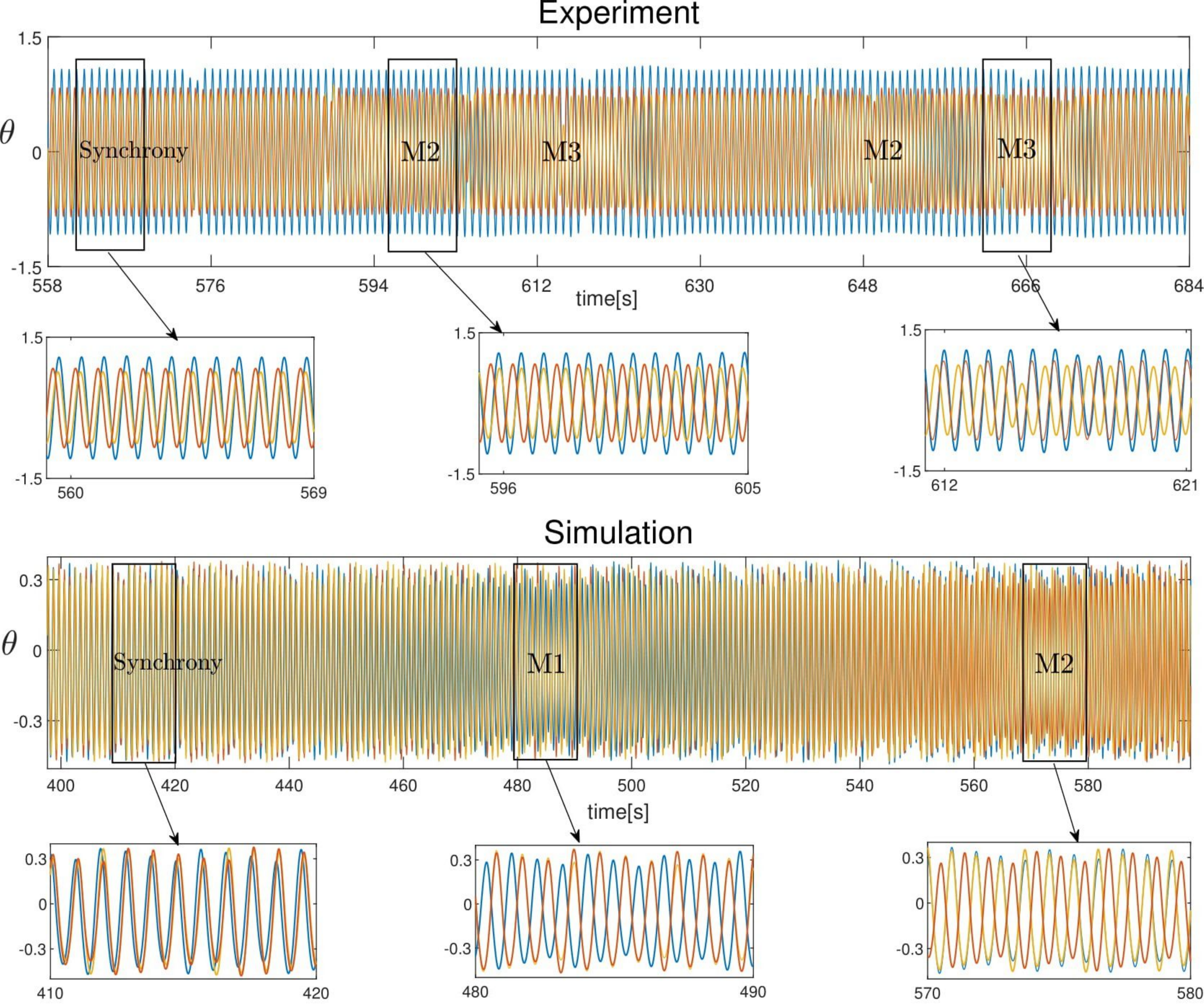}}
\caption{Heteroclinic switchings corresponding to data point $B$ in parameter region Fig.~\ref{fig3} in experiment (up) and simulation (down). The intervals in which metronomes makes a phase slip are labeled with number of the metronome.}
\label{fig4a}
\end{figure}
\begin{figure}
\resizebox{1.\columnwidth}{!}{
 \includegraphics{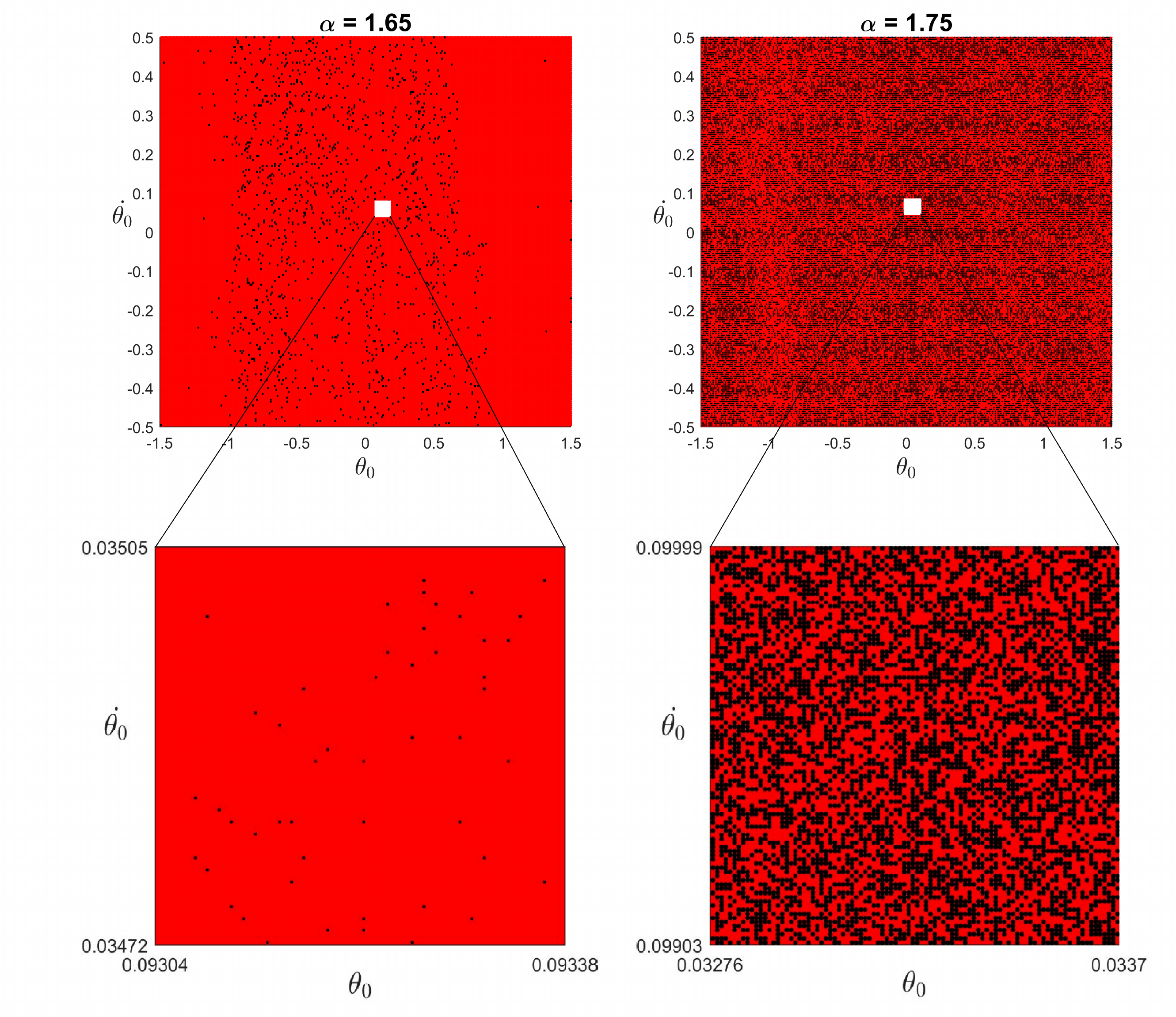}}
\caption{Basins of attraction for chimera states (red) and switchings (black) in parameter points $C (\alpha=1.67)$ and $D (\alpha=1.75)$ in parameter region Fig.~\ref{fig3}. In the chimera region (point C, left panel) the switching behavior is observed in rare black dots only. Close to the stability boundary of the chimera region (point D, right panel), the basin of attraction of the switching behavior has developed riddling structure.}
\label{fig5}
\end{figure}

\begin{figure}
\resizebox{1.\columnwidth}{!}{
 \includegraphics{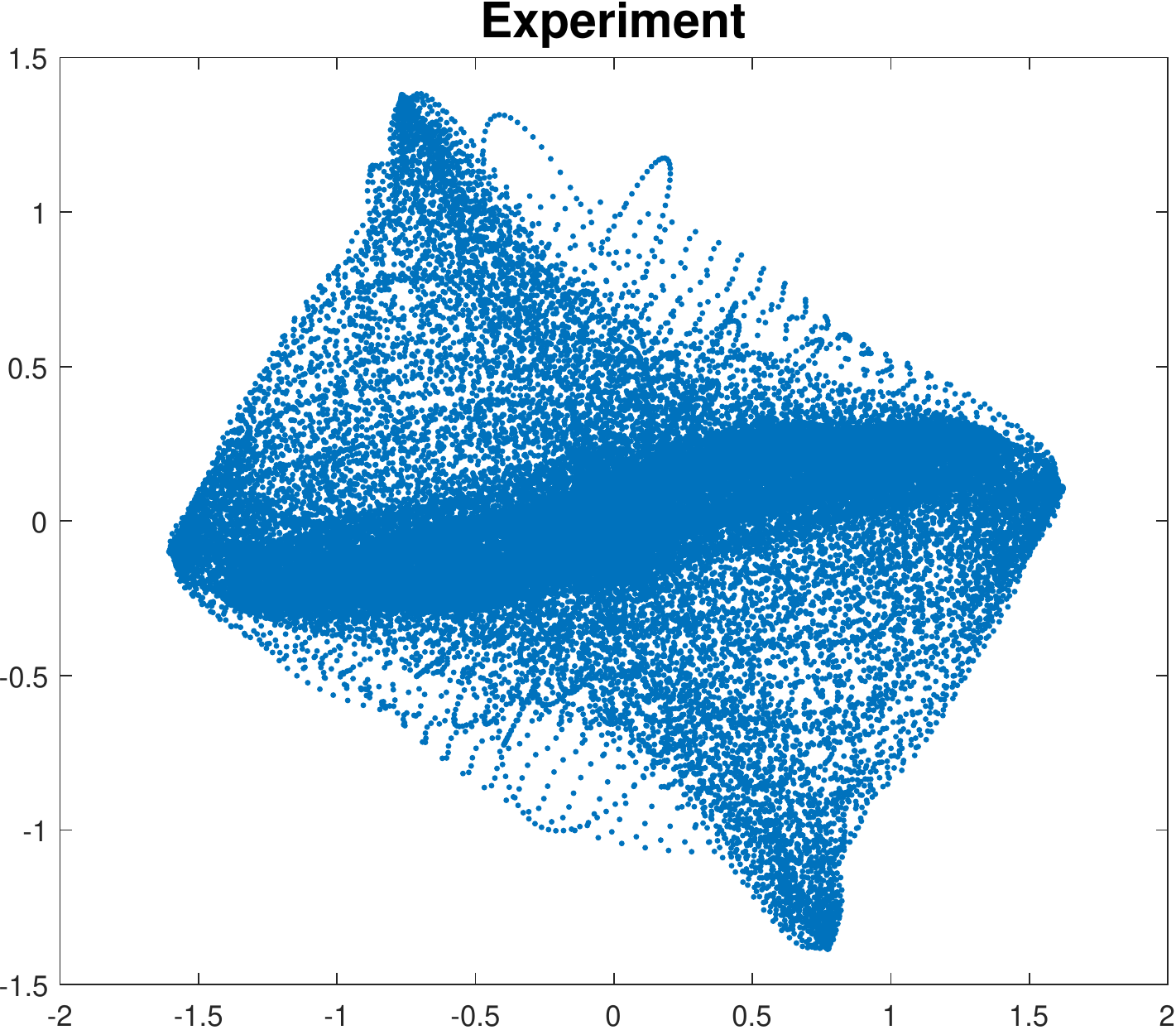} 
 \includegraphics{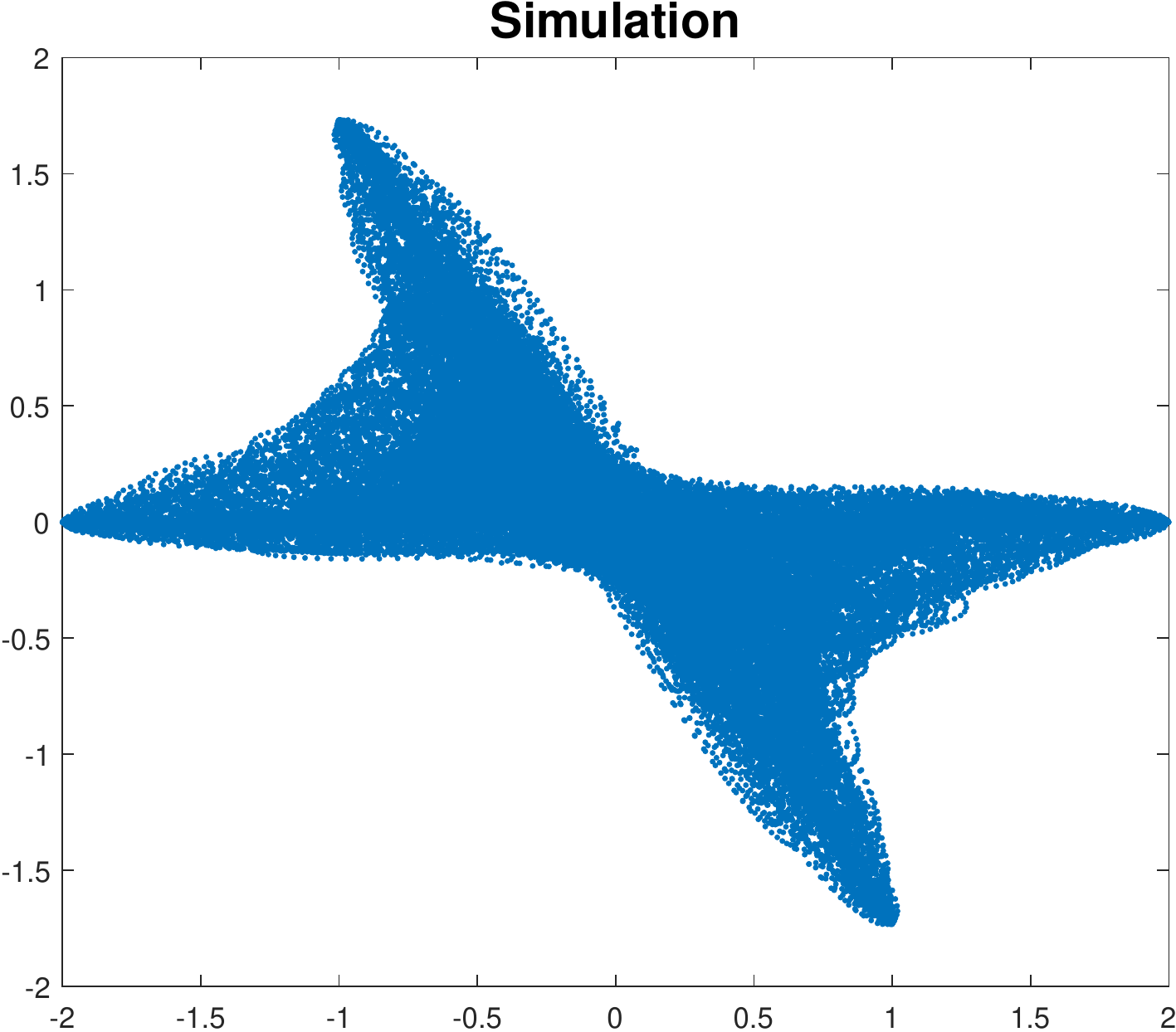} }
\caption{$\zeta$-lattice representation of heteroclinic switching, parameter point $B$. See text.}
\label{fig6}
\end{figure}

Another way of representing the heteroclinic switchings according to Ashwin and Swift \cite{Ashwin_1990}, is to use the equivariant projection function $\zeta = e^{i 2\pi /3}\theta_1 + e^{i 4\pi /3}\theta_2 + \theta_3$ of $T^3$ onto $T^2$ torus. The heteroclinic switching, data point B, is presented in the complex $\zeta$-plane for experiment and simulation, Fig.~\ref{fig6}. Notably, $\zeta$-representation of point $B$ is qualitatively similar to the Fig.9(g) from Ref.~\cite{Ashwin_1990}.
\begin{figure}
\resizebox{1.\columnwidth}{!}{
 \includegraphics{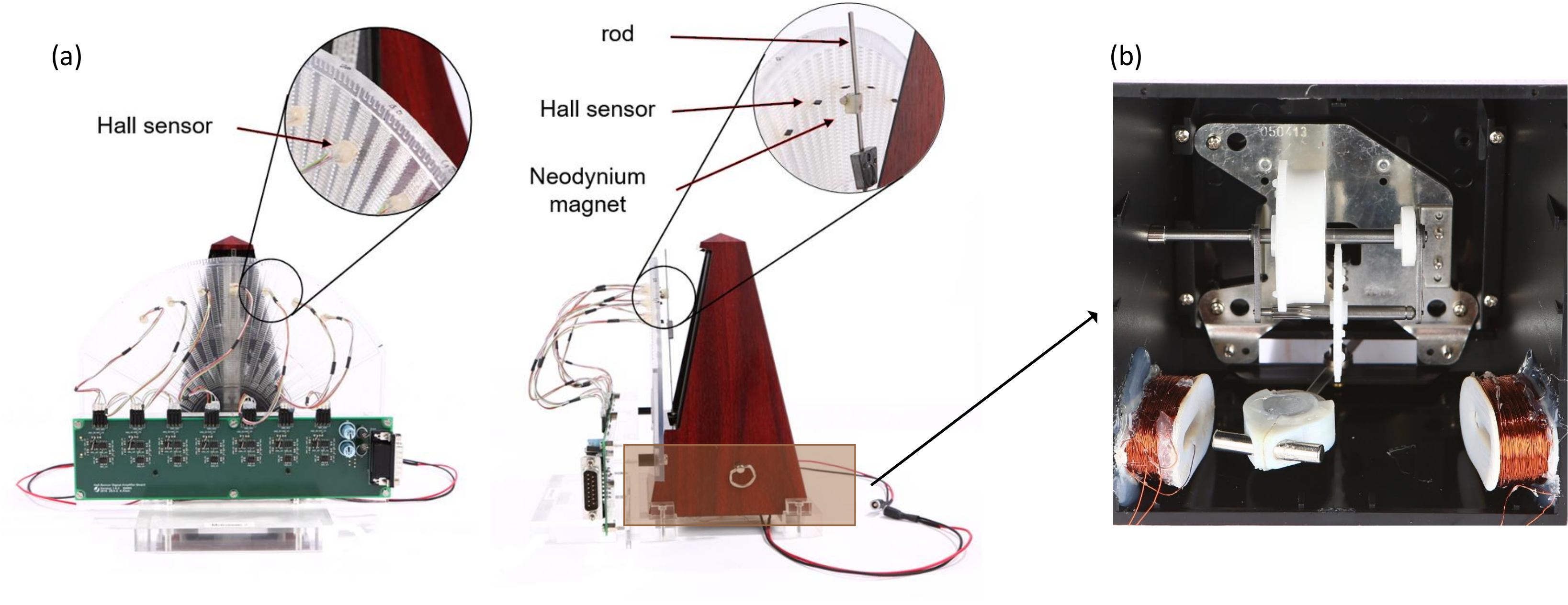}}
\caption{Metronome readout coupling structures. (a) angle readout of metronome using 7 Hall sensors on the sensor holder. (b) coupling implementation mechanisms (bottom view) using coils and magnet. The coupling force is effective as magnet enters coils.}
\label{figA}
\end{figure}
\section{Discussion}
\label{DS}
We have presented experimentally, the existence of chimera states in the minimal network of three mechanical oscillators with phase-lagged all-to-all coupling. The system dynamics is described by the Kuramoto model with inertia. The experimental results are in a good agreement with respective numerical simulations. We observed that chimera states reside in a large area of parameter space between synchronization and rotating waves. By varying the phase-lag parameter $\alpha$ and coupling strength $\mu$, we identified a wide enough layer of switching behaviour at the border of the chimera region. The behaviour there consists in switchings between three saddle chimera states as well as synchronization (for $\alpha<\pi /2$) or rotating waves (for $\alpha>\pi /2$). The behavior there is extremely sensitive to the initial conditions, such that its basin is riddling with fine-grained structure.  

Our experimental setup is flexible regarding varying the coupling configurations and online change of phase-lag and magnitude of coupling. Due to the disposition of the implemented coils, the introduced coupling force has short range characteristic (see appendix for details). In the future perspectives, using a modified coil structure, we expect to achieve effective coupling force in the whole angle range of rod. Moreover, the number $N$ of the metronomes in the setup will be increased. In particular, we plan to analyze the case $N=4$ which is currently under active discussions in the literature as it represents the smallest configuration where two-group dynamics can be resolved. Beyond this, an actual talk is to cope with the effect of non-global coupling topologies having a perspective importance in different fields. 
\section{Appendix}

\subsection{Experiment details}
\label{ExA}
The Hall sensors are mounted on 3D printed plastic sensor holder, whenever the rod Neodynium magnet passes a Hall sensor it causes a zero crossing of its output signal. We use 7 Hall sensors for each metronome, all Hall sensor signals are 12 bit digitized with a sampling rate of 1 kHz by a National Instrument® PCI DAQ card (ADC in Fig.~\ref{fig1}) placed in a standard i7 PC. A Simulink Real-Time® implementation on this PC performs a real-time reconstruction of the rod angles of the three metronomes based on interpolation of lookup tables which are created by an automated calibration procedure. The coupling data are converted at 1 kHz with 12 bit resolution to analog signals by another National Instrument® PCI DAQ card (DAC in Fig.~\ref{fig1}). These analog signals are amplified separately by 6 amplifiers and fed through the actuating coils mounted at the bottom of the metronomes on both sides of each counter weight, Fig.~\ref{figA}b. Neodynium magnets mounted on the counter weights increase the magnetic force on the rods and enable both, repelling and attracting forces. The amplifiers allow for a precise calibration of the coil currents to ensure that equal coupling terms produce equal forces on the rods.
\subsection{Simulation details}
\label{SimA}
Chimera states in Fig.~\ref{fig2} were obtained by solving Eq.~\ref{K1}, using Runge-Kutta 4th order method with step size $h = 0.001$ and initial conditions with format $(\theta, \dot{\theta})$ as follows: $(-1.5, 0.5)$ for $M1$, $(-0.64, 0.3)$ for $M2$, and $(-0.01, 0.2)$ for $M3$. Other parameters of Eq.~(\ref{K1}) are adopted from Ref.~\cite{Wojewoda2016}. The parameter region Fig.~\ref{fig3} was obtained using these fixed initial conditions and parameter step sizes $\delta_{\alpha} = 0.01$ and $ \delta_{\mu} / (\mu) = 0.01$. The basins of attraction for chimera states and switchings in Fig.~\ref{fig5} was obtained with random initial conditions and parameter step sizes $\delta_{\theta} = 0.01$ and $\delta_{\dot{\theta}} = 0.005$. The white samples of basins of attraction for chimera states and switchings in Fig.~\ref{fig5} was obtained with random initial conditions and parameter step sizes 
$\delta_{\theta} = \delta_{\dot{\theta}} = 10 ^ {-5}$.
\bibliographystyle{unsrt}  
\bibliography{references}  

\end{document}